\documentclass{article}
\usepackage{spconf,amsmath,graphicx}
\usepackage{amssymb}
\usepackage{enumitem}
\setlist{nosep, leftmargin=14pt}
\usepackage{subfig}
\usepackage[linesnumbered,ruled,vlined]{algorithm2e}
\usepackage{pifont}
\usepackage{newunicodechar}
\newunicodechar{✓}{\ding{51}}
\newunicodechar{✗}{\ding{55}}
\usepackage{multirow}
\title{Riemannian Prediction of Anatomical Diagnoses in Congenital Heart Disease based on 12-lead ECGs}

\name{Muhammet Alkan$^{\star}$  \qquad Gruschen Veldtman$^{\dagger}$ \qquad Fani Deligianni$^{\star}$}

\address{$^{\star}$ School of Computing Science at University of Glasgow, Glasgow, Scotland, UK \\ $^{\dagger}$ Golden Jubilee National Hospital, Glasgow, Scotland, UK}

\begin{document}
%
\maketitle
\begin{abstract}
Congenital heart disease (CHD) is a relatively rare disease that affects patients at birth and results in extremely heterogeneous anatomical and functional defects. 12-lead ECG signal is routinely collected in CHD patients because it provides significant biomarkers for disease prognosis. However, developing accurate machine learning models is challenging due to the lack of large available datasets. Here, we suggest exploiting the Riemannian geometry of the spatial covariance structure of the ECG signal to improve classification. Firstly, we use covariance augmentation to mix samples across the Riemannian geodesic between corresponding classes. Secondly, we suggest to project the covariance matrices to their respective class Riemannian mean to enhance the quality of feature extraction via tangent space projection. We perform several ablation experiments and demonstrate significant improvement compared to traditional machine learning models and deep learning on ECG time series data.
\end{abstract}
\section{Introduction}
\label{sec:intro}

\begin{figure}[tb]
\begin{minipage}[b]{1.0\linewidth}
  \centering
  \centerline{\includegraphics[width=5.5cm]{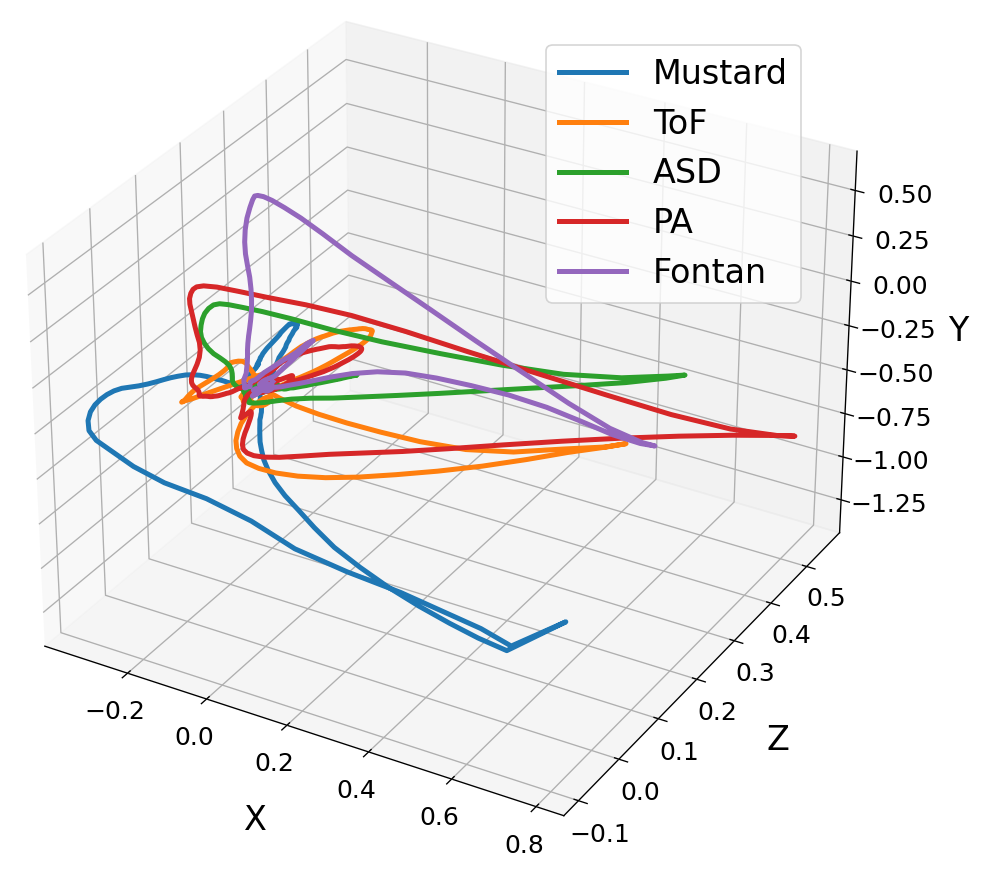}}
\end{minipage}
\caption{Average VCGs across anatomical defects in CHD.}
\label{fig:VCG}
\end{figure}

12-lead ECG is a very common diagnostic and prognostic tool in cardiac diseases. The reason behind this is that it is easy to acquire and it reflects cardiac anatomy and function in high spatio-temporal resolution. In fact, recent work showed that 12-lead ECGs can be converted into 3D representations of the direct cardiac activity during a heart beat based on vectorcardiograms (VCGs) \cite{van2021relation, gopal20213kg}. An example of these representations is shown in Figure \ref{fig:VCG} and it reveals signatures of anatomical defects in Congenital Heart Disease (CHD) based on our data. Several deep learning techniques have been proposed to classify cardiac rhythms and estimate the risk of adverse effects \cite{ribeiro2020automatic}. These methods showed impressive results with large datasets that include millions of patients and ECG recordings. However, it is not clear how they can extend in relatively rare and extremely heterogeneous cases.

In congenital heart disease (CHD) abnormalities in structure and function are present at birth and affect around $1\%$ of babies. In other words, patients are born with genetic defects that differ significantly from the cardiac abnormalities emerging later in life. Therefore, the efficiency of deep learning methods developed on a broader population is limited due to lack of large scale representative data and extreme physiological variations in both anatomy and function.

Inspired by successful work on Riemannian classification \cite{yger2016riemannian}, we proposed to use the covariance structure of the 12-lead ECGs to predict anatomic diagnosis associated with CHD as an initial step toward mortality prediction. 
Our contributions are:
\begin{itemize}
\item We exploit multiple class-dependent tangent spaces to project covariances matrices to their class mean and extract more coherent features for classification. 
\item We exploit a covariance mixing regularisation technique for augmentation. Similar to mixup approach that interpolates samples in a linear way based on a factor $\alpha$, this approach performs the interpolation on a Riemannian manifold with respect to the underlying covariance matrices. 
\item We validate our method in ECGs obtained from patients with congenital heart disease, which represent an example of extremely small and imbalanced dataset. 
\end{itemize}

\section{Methods}
\label{sec:methods}

\subsection{Data}
\label{ssec:data}

\begin{figure}[htb]
\begin{minipage}[b]{1.0\linewidth}
  \centering
  \centerline{\includegraphics[width=7.5cm]{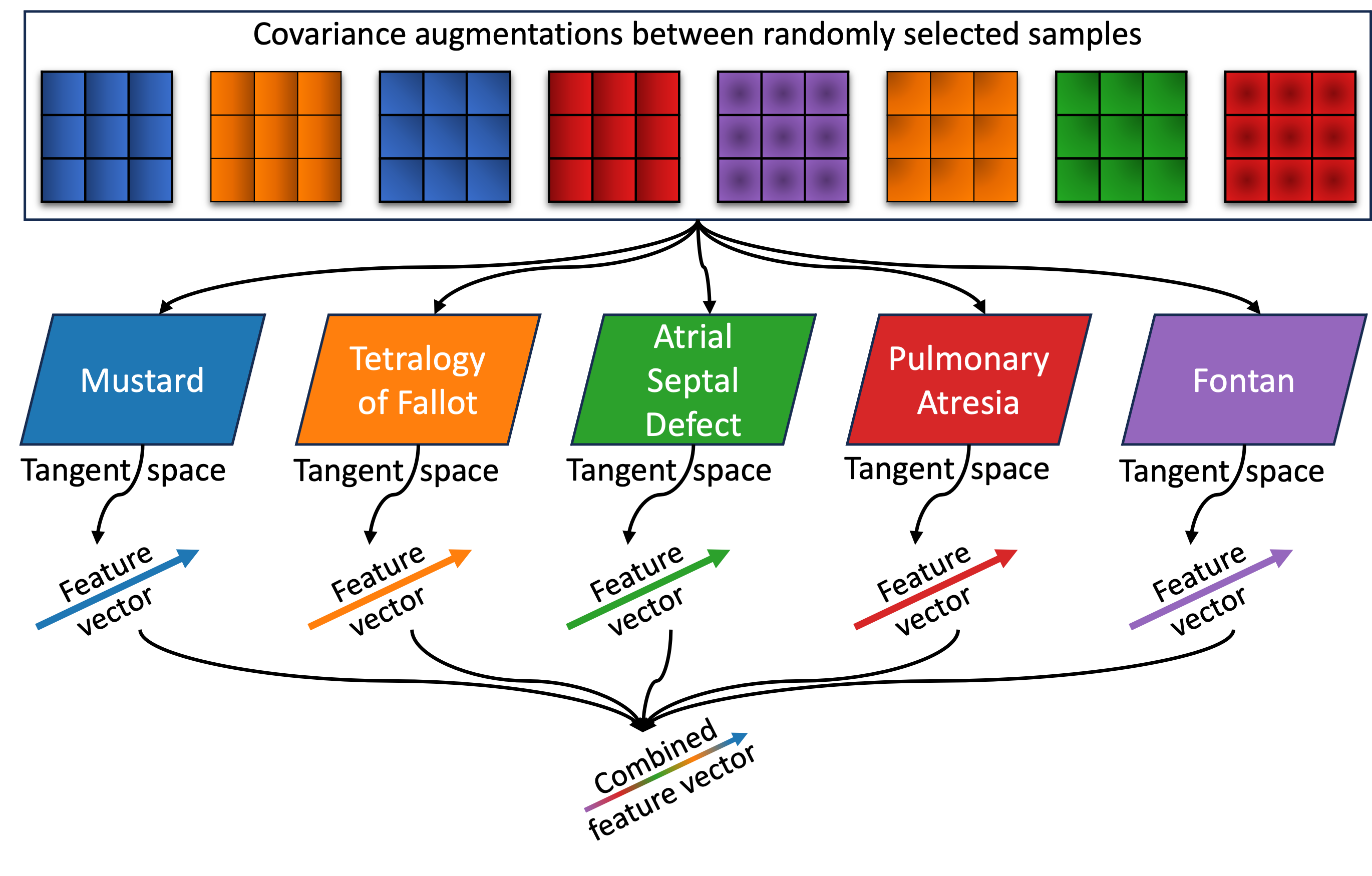}}
\end{minipage}
\caption{Multiple tangent space concept.}
\label{fig:multipleTangent}
\end{figure}

4153 12-lead ECGs were extracted from 436 patients (194 female) with CHD under regular follow-up at the Scottish Adult Congenital Cardiac Service based at the Golden Jubilee National Hospital in Scotland. ECGs in atrial flutter or atrial fibrillation were excluded, as were atrioventricular paced rhythms, as one of our primary aims was to use ECGs in sinus rhythm to predict diagnosis. The most common condition was tetralogy of fallot (ToF, 39.9\%), followed by atrial septal defect (ASD, 17.6\%) and pulmonary atresia (PA, 16.7\%). Mean ECG age was 33 years (SD 11.7 and 75-25\% IQR(40,23)). Patients with no documented ECGs or those in atrial flutter or fibrillation or other heart rhythm abnormalities including being paced, at the time of the ECG were excluded. For patients with more than one anatomic diagnosis, the dominant diagnosis was considered the primary diagnosis. Extracted 436 patients are summarised as follows: 173 patients with ToF, 77 patients with ASD, 73 patients with PA, 66 patients with Fontan and 47 patients with Mustard.


\subsection{From Common to Multiple Tangent Spaces}
\label{sec:models}

ECG signal $Y$ is described as a $12 \times n$ time series data. For Riemannian manifold classifier, the covariance structure of the multichannel (12-lead) ECG signal is estimated and projected into a flat space. This process allows more accurate estimation of linear operations. We applied spatial filtering $F$ to enhance the signal-to-noise ratio and remove the artifacts \cite{rivet2009xdawn}. Then, the covariance matrix is estimated as $\mathbf{C} = F(Y) \cdot F(Y)^\intercal$ which reflects the correlations between each pair of the leads. Since covariance matrices are Symmetric Positive Definite (SPD) matrices, they must be analysed in a Riemannian manifold rather than Euclidean space. Riemannian metric rather than Euclidean metric is used to project covariance matrices onto tangent space while respecting their geometry \cite{barachant2011multiclass}. 
To achieve this, typically, the covariance matrices are projected onto a common tangent space based on Equation \ref{eq:tangentSpace} \cite{yger2016riemannian}. 

\begin{equation}
\label{eq:tangentSpace}
\mathbf{V_i^{C}} = {\rm upper} \left({\bf C}^{-{\frac{1}{2}} } {\rm Log}_{{\bf C}} \left({\bf C}_i \right) {\bf C}^{-{\frac{1}{2}}} \right)
\end{equation}

\begin{algorithm}[htb]
\KwData{12-lead ECG signals}
\KwResult{Anatomic diagnosis associated with CHD}
 initialise SPLO splits and align R Peaks\;
\While{there is a 12-lead ECG}{
  calculate covariance matrices based on the \\spatial filtering $F$\;
  \If{mixup}{
   mixup on covariance matrices using \textbf{Eq. \ref{eq:meanRiemannian}}\;
   }
  \If{projection}{
      \uIf{multiple tangent space}{
      fit a tangent space for each class\;
      project covariance matrices to those tangent spaces using \textbf{Eq. \ref{eq:tangentSpace}}\;
      combine each output to get a feature vector to train on the model
       }
       \Else{
       fit only a single tangent space\;
       project covariance matrices to the \\tangent space using \textbf{Eq. \ref{eq:tangentSpace}}\;
       get feature vector to train on the model
       }
   }
  make classification\;
 }
\caption{\centering Prediction of anatomical diagnoses}
\label{alg:pseudocode}
\end{algorithm}

Where $\bf C_i$ is the covariance matrix to be projected onto the tangent space at point $\bf C$, which represents the Riemannian mean of all the covariance matrices. This projection enhances the performance of classifiers that depend on distance metrics between the sample covariance matrices and it has been successful in processing high-dimensional neurophysiological data \cite{barachant2011multiclass, congedo2017riemannian, barachant2013classification}. However, it assumes that $\bf C$ and $\bf C_i$ are relatively close. 


Here, we hypothesise that projecting each covariance matrix to its corresponding class mean will improve the quality of the mapping, since the distance will be smaller compared to the corresponding distance with the global Riemannian mean. Thus, we fitted a different tangent space for each class and then combined each of the outputs into a new feature vector. After tangent space projection, each covariance matrices is represented as a vector $\mathbf{V}$ of size $n\times(n+1)/2$, where $n$ is the dimension of the covariance matrices. Each covariance matrix is mapped into tangent space by keeping the upper triangular part of the resulting symmetric matrix as denoted in Equation \ref{eq:tangentSpace}.

Also, an illustration of the multiple tangent space concept can be seen in Figure \ref{fig:multipleTangent}. Each output of the tangent spaces can be combined into a single enriched feature vector that is fed as input to the classification model. It also allows us to balance potential data issues by using covariance augmentation, described below, on feature vectors of the underrepresented classes. All the algorithmic steps of our proposed approach are demonstrated in Algorithm \ref{alg:pseudocode}.

\subsection{Augmentation of Covariance Matrices}

Since we have limited datasets, we use a covariance mixing technique to generate more samples in a controlled way similar to \cite{zoumpourlis2022covmix}. To apply mixing, we sample $\alpha$ from a beta distribution on the interval [0, 1], and compute the weighted Riemannian mean according to the Riemannian distance metric between the randomly selected covariance matrices. We also tried to control mixing by restricting the range of sampled values for $\alpha$, but the best results were obtained with no restrictions. Riemannian mean that minimises the sum of squared Riemannian distances to the given two SPD matrices was calculated to find the weighted Riemannian mean as in Equation \ref{eq:meanRiemannian}, where $w_i$ represents a weight matrix generated using the $\alpha$ value, $d_R$ represents the Riemannian distances to the SPD matrices.


\vspace{-0.25\baselineskip}

\begin{equation}
\label{eq:meanRiemannian}
\mathbf{C_{aug}} = \arg \min_{\mathbf{C}} \sum_i w_i \ d_R (\mathbf{C}, \mathbf{C}_i)^2
\end{equation}

\begin{figure}[htb]
\centering
\begin{minipage}[b]{.49\linewidth}
\centering
  \subfloat[\centering t-SNE on tangent space\label{fig:TSNE1}]
      {\includegraphics[width=4cm]{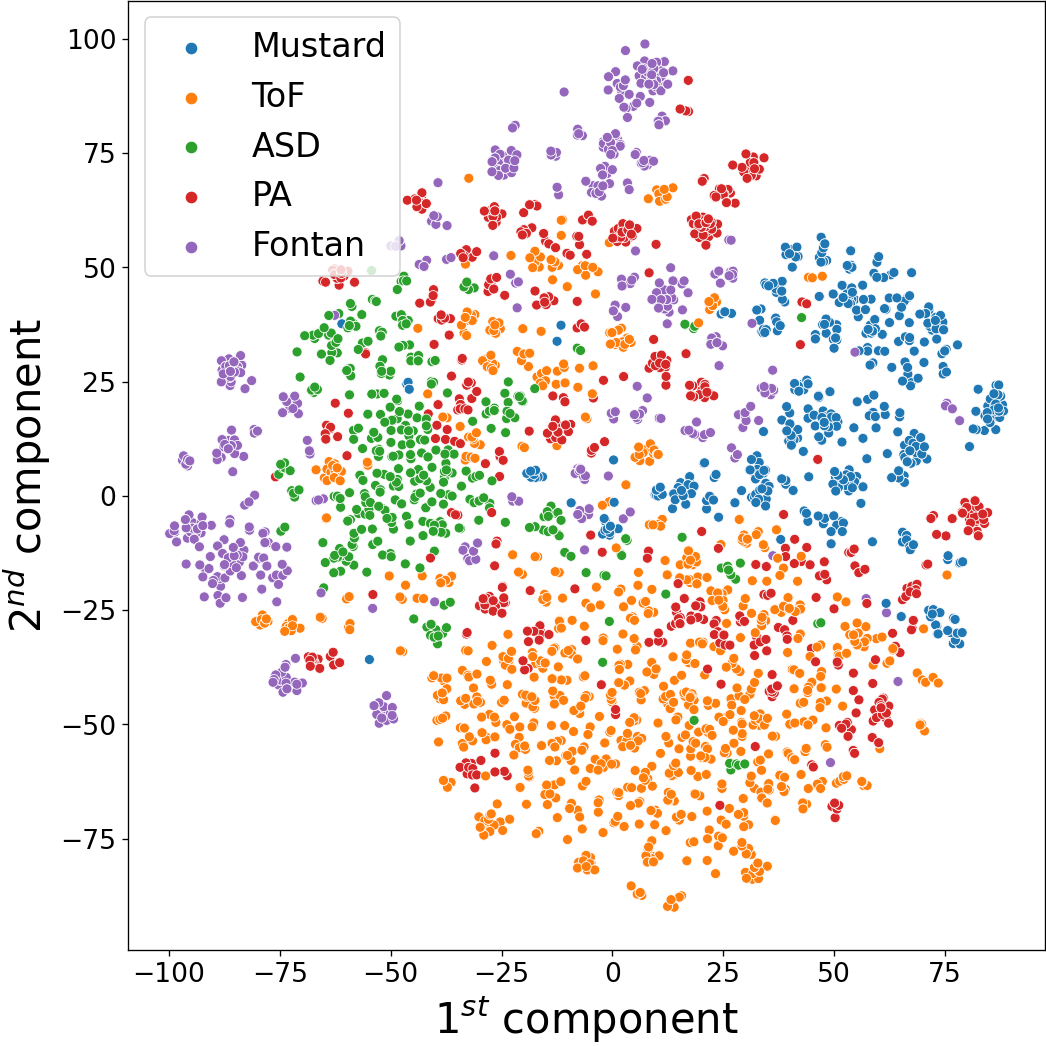}}
\end{minipage}
\begin{minipage}[b]{0.49\linewidth}
\centering
  \subfloat[\centering t-SNE with covariance
augmentations on tangent space\label{fig:TSNE2}]
      {\includegraphics[width=4cm]{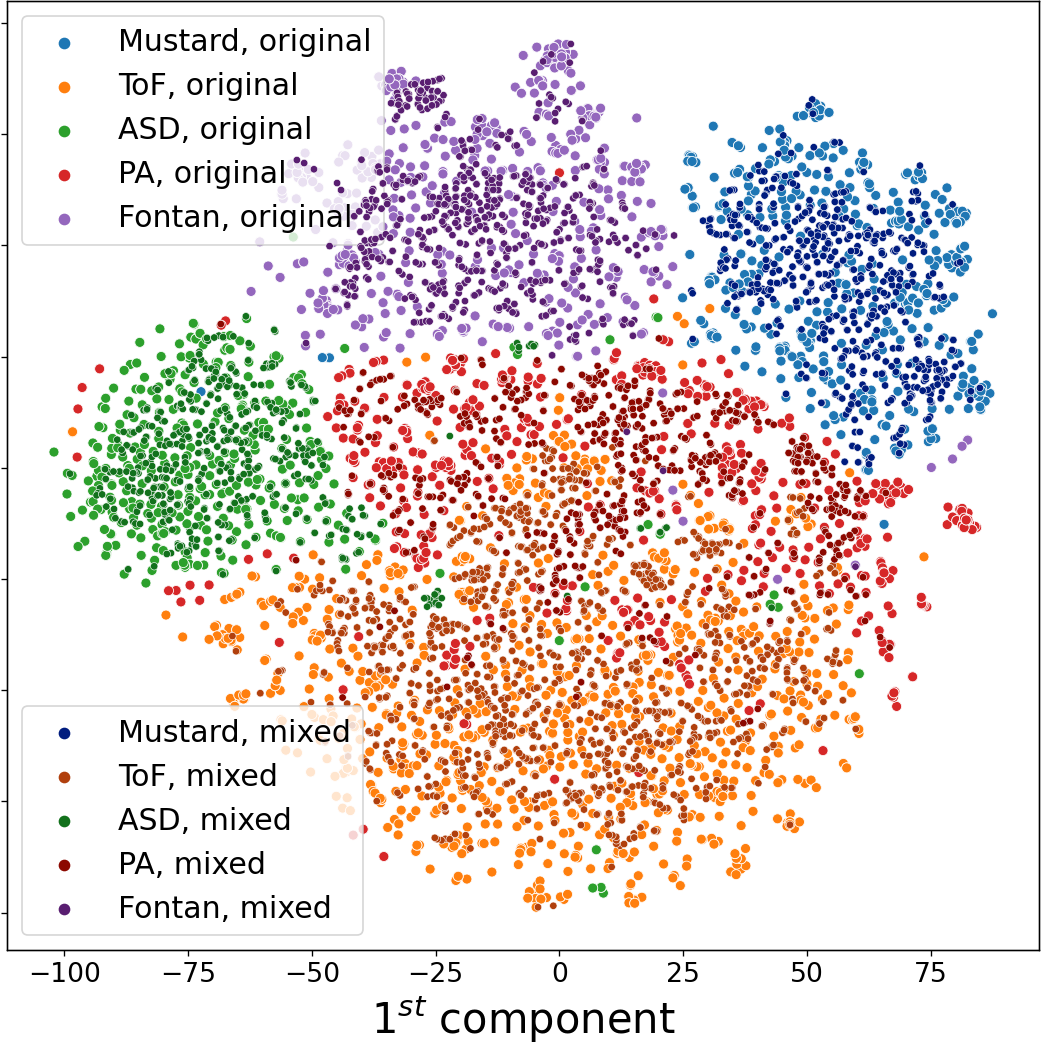}}
\end{minipage}
\caption{\centering t-SNE visualizations on the tangent space.}
\label{fig:TSNE}
\end{figure}

Instead of mixing all the data we have, we tried to focus only on the classes that are not easily distinguishable by the model. t-SNE visualisations \cite{van2008visualizing} on the tangent space using only original data, and mixed data combined with the original are shown in Figures \ref{fig:TSNE1} and \ref{fig:TSNE2}, respectively.

\vspace{-1\baselineskip}

\section{Results}
\label{sec:results}

12-lead ECG data were extracted from ECG PDF documents obtained via the Marquette™ 12SL by GE Healthcare analysis program. Subsequently, ECG data were digitised and pre-processed to align the R peak points across heart beats for all patients. The average length of raw ECGs is 2.5 seconds and R peak aligned ECGs is around 1 second, and they are sampled at a rate of 500 samples per second. Aligned ECG data gave better results than raw ECG data, regardless of input or model type.

Training-testing split of data was repeated 100 times based on pseudo-randomised, stratified patient leave-out (SPLO) evaluation to ensure that the testing set of patients was representative of all the classes. One patient for each class is randomly selected for testing. All of testing patients'  ECG data are removed from training and only data from the rest of patients are used in the training. This training procedure is repeated 100 times in a pseudo-randomised manner and the average performance results (accuracy, AUC and F1 macro) along with standard deviation are reported. On average, each patient has 10 ECG recordings (for a very small number of patients this number can vary from 2 to 40). Firstly, the data are split into different sets and subsequently further processing, such as covariance augmentations, is performed only on the training data.

Three different classifiers were used to compare our feature extraction framework: Support Vector Machine (SVM), Minimum Distance to Means (MDM), and Multilayer Perceptron (MLP). We have done ablation studies with MLP model as well with baseline models SVM and MDM. MDM performs classification by the nearest centroid. A centroid of the covariance matrices is estimated for each of the classes and then, for each new covariance sample, the class is estimated according to the nearest centroid. MLP model was trained with the different feature extraction frameworks as mentioned in Table \ref{tab:ablation}. Table \ref{tab:ablation} shows the different strategies applied for augmentation and covariance projection. $(DEF)$ denotes application of the classification directly on the R peak aligned 12-lead ECG data as the default setting. ($VCG$) represents augmentations based on ECG vectorcardiogram space. 
Dower transformation was used to project 12-lead ECG data into a 3-dimensional VCG space. Subsequently, we applied rotations along all three orthogonal axes from 5 to 45 degrees and projected the augmented VCG back into 12-lead ECG space \cite{gopal20213kg}. ($VCG$) augmentation performed poorly in classifying covariance matrices $(VCG\_COV)$.
In the next ablation steps, $(COV)$ represents that covariance matrices were used. The final step was the projection of these matrices into tangent space (single: $TS$ or multiple: $MTS$). The best results were obtained using multiple tangent spaces with covariance augmentations $(MTS\_COV)$.

\renewcommand{\arraystretch}{1.2}
\begin{table}[htb]
\centering
\caption{Ablation study: AUC scores for different methods.}
\label{tab:ablation}
\resizebox{\columnwidth}{!}{
\begin{tabular}{|c|cccc|c|c|c|}
\hline
\multirow{2}{*}[-8pt]{\large\textbf{Name}} & \multicolumn{4}{c|}{\textbf{Feature Extraction Framework}} & \multirow{2}{*}[-8pt]{\textbf{\begin{tabular}[c]{@{}c@{}}AUC\\ (MLP)\end{tabular}}} & \multirow{2}{*}[-8pt]{\textbf{\begin{tabular}[c]{@{}c@{}}AUC\\ (SVM)\end{tabular}}} & \multirow{2}{*}[-8pt]{\textbf{\begin{tabular}[c]{@{}c@{}}AUC\\ (MDM)\end{tabular}}} \\ \cline{2-5}
 & \multicolumn{1}{c|}{\begin{tabular}[c]{@{}c@{}}VCG\\ Augmentations\end{tabular}} & \multicolumn{1}{c|}{\begin{tabular}[c]{@{}c@{}}Covariance\\ Matrix\end{tabular}} & \multicolumn{1}{c|}{\begin{tabular}[c]{@{}c@{}}Covariance\\ Augmentations\end{tabular}} & \begin{tabular}[c]{@{}c@{}}Tangent\\ Space\end{tabular} &  &  &  \\ \hline
$DEF$ & \multicolumn{1}{c|}{✗} & \multicolumn{1}{c|}{} & \multicolumn{1}{c|}{} &  & 0.73 ± 0.05 & 0.80 ± 0.08 & NA \\ \hline
$VCG$ & \multicolumn{1}{c|}{✓} & \multicolumn{1}{c|}{} & \multicolumn{1}{c|}{} &  & 0.76 ± 0.08 & 0.59 ± 0.04 & NA \\ \hline
$COV$ & \multicolumn{1}{c|}{✗} & \multicolumn{1}{c|}{✓} & \multicolumn{1}{c|}{} &  & 0.77 ± 0.08 & 0.50 ± 0.09 & 0.76 ± 0.09 \\ \hline
$VCG\_COV$ & \multicolumn{1}{c|}{✓} & \multicolumn{1}{c|}{✓} & \multicolumn{1}{c|}{} &  & 0.50 ± 0.00 & 0.50 ± 0.00 & 0.50 ± 0.00 \\ \hline
$TS$ & \multicolumn{1}{c|}{✗} & \multicolumn{1}{c|}{✓} & \multicolumn{1}{c|}{✗} & Single & 0.82 ± 0.08 & 0.72 ± 0.07 & 0.80 ± 0.07 \\ \hline
$TS\_COV$ & \multicolumn{1}{c|}{✗} & \multicolumn{1}{c|}{✓} & \multicolumn{1}{c|}{✓} & Single & 0.83 ± 0.07 & 0.77 ± 0.06 & 0.82 ± 0.06 \\ \hline
$MTS$ & \multicolumn{1}{c|}{✗} & \multicolumn{1}{c|}{✓} & \multicolumn{1}{c|}{✗} & Multiple & 0.81 ± 0.09 & 0.78 ± 0.08 & NA \\ \hline
$\textbf{\textit{MTS\_COV}}$ & \multicolumn{1}{c|}{✗} & \multicolumn{1}{c|}{✓} & \multicolumn{1}{c|}{✓} & Multiple & \textbf{0.84 ± 0.06} & \textbf{0.82 ± 0.08} & NA \\ \hline
\end{tabular}}
\end{table}

\renewcommand{\arraystretch}{1.2}
\begin{table}[htb]
\centering
\caption{\centering Performance metrics of the machine learning models. With \textbf{bold} we highlight the best six models.}
\label{tab:resultsTable}
\resizebox{\columnwidth}{!}{
\begin{tabular}{c|c|c|c|}
\cline{2-4}
 & \textbf{Accuracy} & \textbf{AUC} & \textbf{F1 macro} \\ \hline
\multicolumn{1}{|c|}{\begin{tabular}[c]{@{}c@{}}\textbf{MLP} $(MTS\_COV)$\end{tabular}} & \textbf{0.71 ± 0.10} & \textbf{0.84 ± 0.06} & \textbf{0.69 ± 0.13} \\ \hline
\multicolumn{1}{|c|}{\begin{tabular}[c]{@{}c@{}}\textbf{MLP} $(MTS)$\end{tabular}} & \multicolumn{1}{c|}{\textbf{0.64 ± 0.17}} & \multicolumn{1}{c|}{\textbf{0.81 ± 0.09}} & \multicolumn{1}{c|}{\textbf{0.63 ± 0.18}} \\ \hline
\multicolumn{1}{|c|}{\begin{tabular}[c]{@{}c@{}}\textbf{MLP} $(TS\_COV)$\end{tabular}} & \multicolumn{1}{c|}{\textbf{0.69 ± 0.12}} & \multicolumn{1}{c|}{\textbf{0.83 ± 0.07}} & \multicolumn{1}{c|}{\textbf{0.65 ± 0.13}} \\ \hline
\multicolumn{1}{|c|}{\begin{tabular}[c]{@{}c@{}}\textbf{MLP} $(TS)$\end{tabular}} & \textbf{0.66 ± 0.15} & \textbf{0.82 ± 0.08} & \textbf{0.63 ± 0.17} \\ \hline
\multicolumn{1}{|c|}{\begin{tabular}[c]{@{}c@{}}\textbf{MLP} $(VCG\_COV)$\end{tabular}} & 0.17 ± 0.06 & 0.50 ± 0.00 & 0.05 ± 0.01 \\ \hline
\multicolumn{1}{|c|}{\begin{tabular}[c]{@{}c@{}}\textbf{MLP} $(VCG)$\end{tabular}} & 0.62 ± 0.13 & 0.76 ± 0.08 & 0.54 ± 0.15 \\ \hline
\multicolumn{1}{|c|}{\begin{tabular}[c]{@{}c@{}}\textbf{MLP} $(DEF)$\end{tabular}} & 0.55 ± 0.12 & 0.73 ± 0.05 & 0.48 ± 0.09 \\ \hline
\multicolumn{1}{|c|}{\begin{tabular}[c]{@{}c@{}}\textbf{SVM} $(MTS\_COV)$\end{tabular}} & \textbf{0.66 ± 0.17} & \textbf{0.82 ± 0.08} & \textbf{0.65 ± 0.17} \\ \hline
\multicolumn{1}{|c|}{\begin{tabular}[c]{@{}c@{}}\textbf{SVM} $(MTS)$\end{tabular}} & 0.61 ± 0.13 & 0.78 ± 0.08 & 0.56 ± 0.16 \\ \hline
\multicolumn{1}{|c|}{\begin{tabular}[c]{@{}c@{}}\textbf{SVM} $(TS\_COV)$\end{tabular}} & 0.59 ± 0.10 & 0.77 ± 0.06 & 0.55 ± 0.11 \\ \hline
\multicolumn{1}{|c|}{\begin{tabular}[c]{@{}c@{}}\textbf{SVM} $(TS)$\end{tabular}} & 0.56 ± 0.12 & 0.72 ± 0.07 & 0.53 ± 0.12 \\ \hline
\multicolumn{1}{|c|}{\begin{tabular}[c]{@{}c@{}}\textbf{SVM} $(VCG\_COV)$\end{tabular}} & 0.17 ± 0.06 & 0.50 ± 0.00 & 0.05 ± 0.01 \\ \hline
\multicolumn{1}{|c|}{\begin{tabular}[c]{@{}c@{}}\textbf{SVM} $(DEF_{2D})$\end{tabular}} & 0.65 ± 0.14 & 0.80 ± 0.08 & 0.62 ± 0.15 \\ \hline
\multicolumn{1}{|c|}{\begin{tabular}[c]{@{}c@{}}\textbf{MDM} $(TS\_COV)$\end{tabular}} & \textbf{0.67 ± 0.14} & \textbf{0.82 ± 0.06} & \textbf{0.66 ± 0.11} \\ \hline
\multicolumn{1}{|c|}{\begin{tabular}[c]{@{}c@{}}\textbf{MDM} $(TS)$\end{tabular}} & 0.65 ± 0.15 & 0.80 ± 0.07 & 0.63 ± 0.13 \\ \hline
\multicolumn{1}{|c|}{\begin{tabular}[c]{@{}c@{}}\textbf{MDM} $(VCG\_COV)$\end{tabular}} & 0.17 ± 0.06 & 0.50 ± 0.00 & 0.05 ± 0.01 \\ \hline
\multicolumn{1}{|c|}{\begin{tabular}[c]{@{}c@{}}\textbf{MDM} $(COV)$\end{tabular}} & 0.63 ± 0.15 & 0.76 ± 0.09 & 0.56 ± 0.17 \\ \hline
\end{tabular}}
\end{table}

Table \ref{tab:resultsTable} provides a comparison on the performance metrics (Accuracy, AUC, F1 macro) of the approaches described in ablation Table \ref{tab:ablation} as well as a baseline model with SVM applied on the aligned ECG time-series data $DEF_{2D}$. Using tangent space projection on MLP resulted in a 9\% increase in the AUC score and an 11\% increase in the accuracy. Using covariance augmentations on the single tangent space also yielded a further increase of 1\% in the AUC score and a 3\% increase in the accuracy. Taking a step further and using multiple tangent space projections with covariance augmentations resulted in an 11\% increase in the AUC score and a 16\% increase in the accuracy. Better results were obtained by projecting the augmented covariance matrices into multiple tangent space while respecting their underlying geometry.

Both SVM and MLP models were trained on R peak aligned ECG data $(DEF)$, but for the SVM model 3D ECG data were reshaped to 2D $(DEF_{2D})$. SVM model achieved 7\% better results at this stage for the AUC score. Also, some 3D rotations on the ECG data were tried by projecting it to the VCG space with the help of Dower transformations. Only MLP model was trained on these 3D augmented ECG data $(VCG)$ and achieved 3\% better results for the AUC score. But, they caused very poor results with the next steps that include covariance matrices $(VCG\_COV)$. VCG means without any augmentations for each class can be seen in Figure \ref{fig:VCG}.

\begin{figure}[htb]
\begin{minipage}[b]{1.0\linewidth}
  \centering
  \centerline{\includegraphics[width=8.5cm]{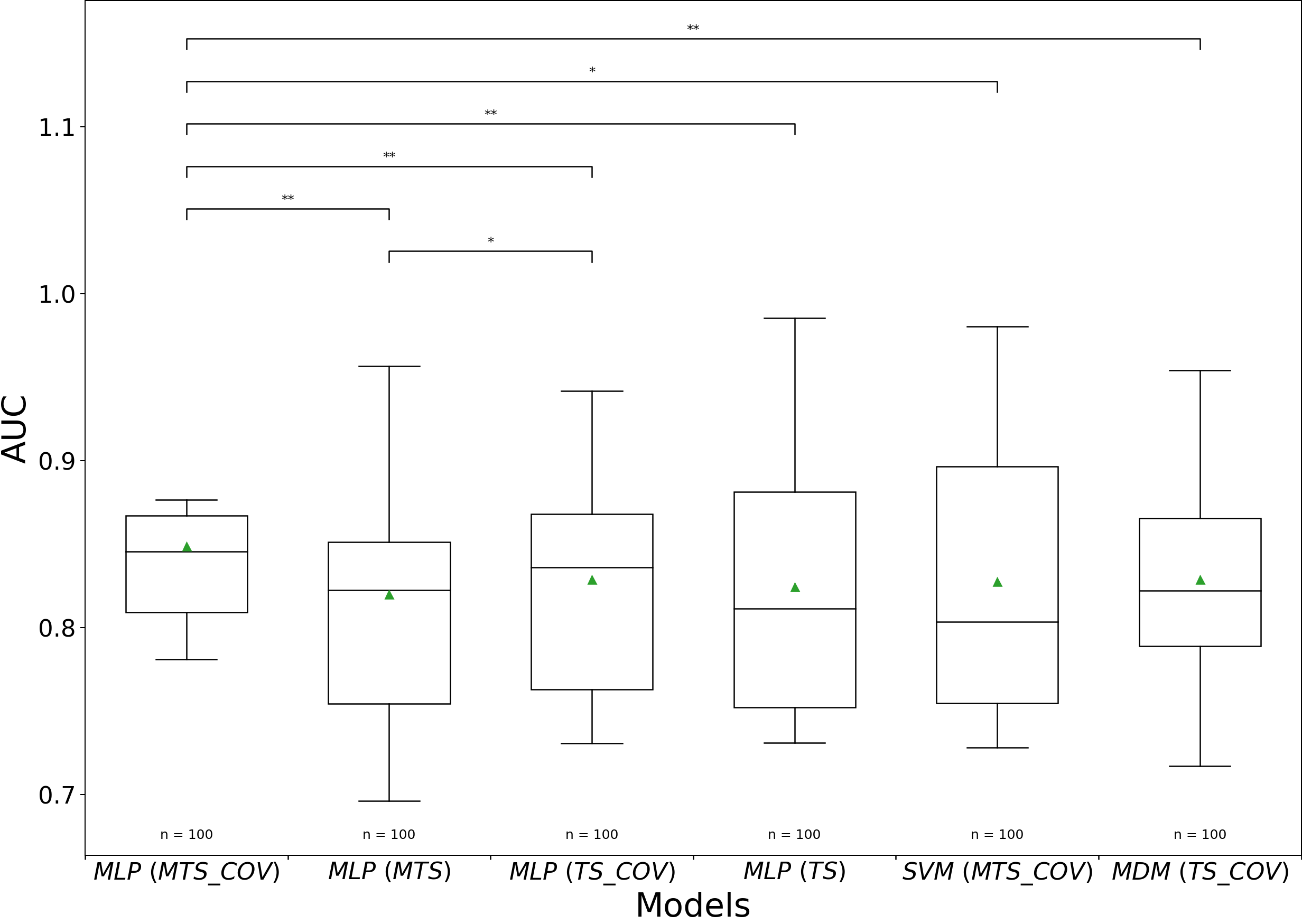}}
\end{minipage}
\caption{AUC scores for the best six approaches.}
\label{fig:AUC}
\end{figure}

Figure \ref{fig:AUC} shows how the top six models compare statistically where statistical significance is based on corrected paired t-test \cite{nadeau1999inference}. All the models except SVM achieved better results on the tangent space and peaked at an AUC score of 82\%. 
There was a statistically significant difference between the AUC results of the $MLP (MTS\_COV)$ model compared to the other models in Figure \ref{fig:AUC}. As shown in Table \ref{tab:resultsTable}, using multiple tangent spaces provided an improvement of 11\%  for the MLP model and 2\% for the SVM model in the AUC score.

\begin{figure}[htb]
\centering
\begin{minipage}[b]{.3\linewidth}
  \subfloat[\centering MLP\label{fig:resD}]
      {\includegraphics[width=2.8cm]{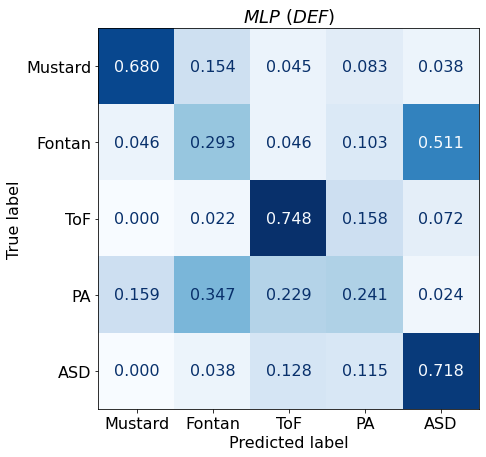}}
\end{minipage}
\hspace{0.5cm}
\begin{minipage}[b]{0.3\linewidth}
  \subfloat[\centering MLP on tangent \newline space\label{fig:resC}]
      {\includegraphics[width=2.3cm]{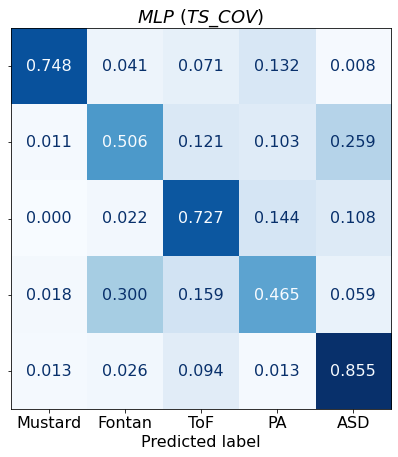}}
\end{minipage}
\hspace{0.01cm}
\begin{minipage}[b]{.3\linewidth}
  \subfloat[\centering MLP on multiple \newline tangent space\label{fig:resB}]
      {\includegraphics[width=2.7cm]{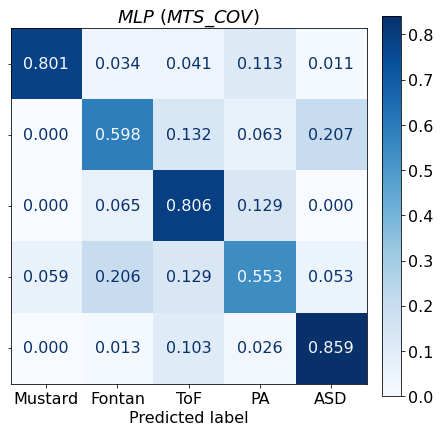}}
\end{minipage}
\caption{\centering Confusion matrices of $DEF$, $TS\_COV$ and $MTS\_COV$ approaches with an MLP classifier.}
\label{fig:conMat}
\end{figure}

\begin{figure}[h!] 
\begin{minipage}[b]{1.0\linewidth}
\centering
\centerline{\includegraphics[width=4.5cm]{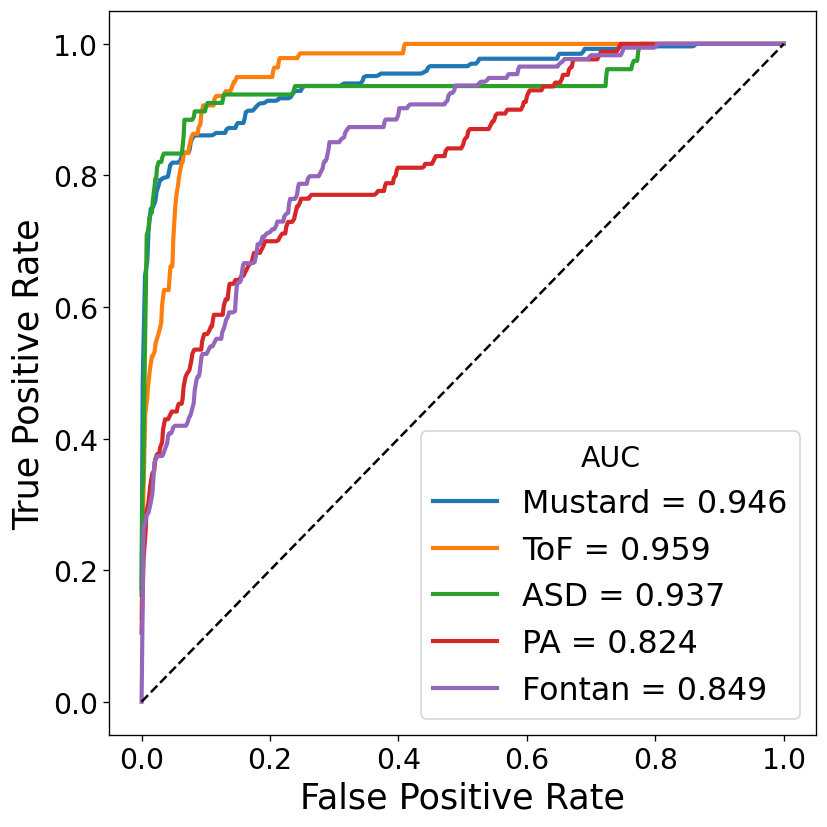}}
\end{minipage}
\caption{AUC for the best model ($MTS\_COV$).}
\label{fig:AUC_best}
\end{figure}

There was a statistically significant difference between the AUC results of the $MLP (MTS\_COV)$ model compared to the other models in Figure \ref{fig:AUC}. 
The confusion matrices in Figure \ref{fig:resD}, \ref{fig:resC} and \ref{fig:resB}, reflect the improvement in performance with the application of the multiple tangent space projection and covariance augmentation. Furthermore, Figure \ref{fig:AUC_best} shows AUC score for each class on the best model $MTS\_COV$, evaluated using One-vs-Rest (OvR) strategy.

\vspace{-0.8\baselineskip}

\vspace{-1\baselineskip}

\section{Conclusions}
\label{sec:conclusions}

We demonstrated promising results on 12-lead ECG classification of anatomical diagnosis in congenital heart disease. Our proposed projection of the augmented covariance matrices to multiple Riemannian spaces yields significantly better results in improving classification performance with small and extremely imbalanced 12-lead ECG data.

\section{Compliance with Ethical Standards}
This study was performed in line with the principles of the Declaration of Helsinki. Study approval was obtained from the Institutional Governance Division of the NHS Golden Jubilee National Hospital.
This work was supported by the School of Computing Science at University of Glasgow and Golden Jubilee National Hospital leadership and governance teams. F.D. is supported by funding from grant EP/W01212X/1. None of the authors have anything to disclose relevant to this work. We will make our code publicly available to enable reproducibility.\\

\bibliographystyle{IEEEbib}
\bibliography{refs}

\begin{thebibliography}{10}

\bibitem{van2021relation}
Peter~M van Dam, Machteld Boonstra, Emanuela~T Locati, and Peter Loh,
\newblock ``The relation of 12 lead ecg to the cardiac anatomy: the normal cineecg,''
\newblock {\em Journal of Electrocardiology}, vol. 69, pp. 67--74, 2021.

\bibitem{gopal20213kg}
Bryan Gopal, Ryan Han, Gautham Raghupathi, Andrew Ng, Geoff Tison, and Pranav Rajpurkar,
\newblock ``3kg: Contrastive learning of 12-lead electrocardiograms using physiologically-inspired augmentations,''
\newblock in {\em Machine Learning for Health}. PMLR, 2021, pp. 156--167.

\bibitem{ribeiro2020automatic}
Ant{\^o}nio~H Ribeiro, Manoel~Horta Ribeiro, Gabriela~MM Paix{\~a}o, Derick~M Oliveira, Paulo~R Gomes, J{\'e}ssica~A Canazart, Milton~PS Ferreira, Carl~R Andersson, Peter~W Macfarlane, Wagner Meira~Jr, et~al.,
\newblock ``Automatic diagnosis of the 12-lead ecg using a deep neural network,''
\newblock {\em Nature communications}, vol. 11, no. 1, pp. 1760, 2020.

\bibitem{yger2016riemannian}
Florian Yger, Maxime Berar, and Fabien Lotte,
\newblock ``Riemannian approaches in brain-computer interfaces: a review,''
\newblock {\em IEEE Transactions on Neural Systems and Rehabilitation Engineering}, vol. 25, no. 10, pp. 1753--1762, 2016.

\bibitem{rivet2009xdawn}
Bertrand Rivet, Antoine Souloumiac, Virginie Attina, and Guillaume Gibert,
\newblock ``xdawn algorithm to enhance evoked potentials: application to brain--computer interface,''
\newblock {\em IEEE Transactions on Biomedical Engineering}, vol. 56, no. 8, pp. 2035--2043, 2009.

\bibitem{barachant2011multiclass}
Alexandre Barachant, St{\'e}phane Bonnet, Marco Congedo, and Christian Jutten,
\newblock ``Multiclass brain--computer interface classification by riemannian geometry,''
\newblock {\em IEEE Transactions on Biomedical Engineering}, vol. 59, no. 4, pp. 920--928, 2011.

\bibitem{congedo2017riemannian}
Marco Congedo, Alexandre Barachant, and Rajendra Bhatia,
\newblock ``Riemannian geometry for eeg-based brain-computer interfaces; a primer and a review,''
\newblock {\em Brain-Computer Interfaces}, vol. 4, no. 3, pp. 155--174, 2017.

\bibitem{barachant2013classification}
Alexandre Barachant, St{\'e}phane Bonnet, Marco Congedo, and Christian Jutten,
\newblock ``Classification of covariance matrices using a riemannian-based kernel for bci applications,''
\newblock {\em Neurocomputing}, vol. 112, pp. 172--178, 2013.

\bibitem{zoumpourlis2022covmix}
Georgios Zoumpourlis and Ioannis Patras,
\newblock ``Covmix: Covariance mixing regularization for motor imagery decoding,''
\newblock in {\em 2022 10th International Winter Conference on Brain-Computer Interface (BCI)}. IEEE, 2022, pp. 1--7.

\bibitem{van2008visualizing}
Laurens Van~der Maaten and Geoffrey Hinton,
\newblock ``Visualizing data using t-sne.,''
\newblock {\em Journal of machine learning research}, vol. 9, no. 11, 2008.

\bibitem{nadeau1999inference}
Claude Nadeau and Yoshua Bengio,
\newblock ``Inference for the generalization error,''
\newblock {\em Advances in neural information processing systems}, vol. 12, 1999.

\end{thebibliography}

\end{document}